    \documentstyle[proceedings,numreferences]{maxent95}  
\begin{opening}
\title{BAYESIAN REASONING VERSUS CONVENTIONAL STATISTICS\protect\\
       IN HIGH ENERGY PHYSICS}
\author{G. D'AGOSTINI}
\institute{Dipartimento di Fisica dell'Universit\`a ``La Sapienza''\\
           Piazzale Aldo Moro 2, I-00185 Roma 
           (Italy)\footnote{Email: dagostini@roma1.infn.it.\  
  URL: http://www-zeus.roma1.infn.it/$^\sim$agostini/
\\ Invited talk at the XVIII International Workshop on 
 Maximum Entropy and Bayesian Methods (Maxent98), Garching/M\"unchen
 (Germany), July 27-31 1998. 
}}
\end{opening}

\runningtitle{BAYESIAN REASONING VS CONVENTIONAL STATISTICS IN HEP}

\begin{document}

\begin{abstract}
The intuitive reasoning of physicists in conditions of uncertainty
is closer to the Bayesian approach than to the frequentist 
ideas  taught at University and which are considered the 
reference framework for handling statistical problems. The combination
of intuition and conventional statistics allows practitioners
to get results which are very close, both in meaning and in numerical 
value, to those obtainable by Bayesian methods, at least in simple
routine applications. There are, however, cases in which  
``arbitrary'' probability inversions produce unacceptable or 
misleading results and in these cases 
the conscious application of Bayesian reasoning 
becomes crucial. Starting from these considerations, I will finally 
comment on the often debated question:
``is there any chance that all physicists will become Bayesian?''
\end{abstract}

% The \begin{document} command comes after the \end{opening}
% command.
%  \begin{abstract}
%    The text of the abstract goes here.
%  \end{abstract}

\section{Introduction}
High Energy Physics
%\footnote{I consider essentially HEP physics, 
%expecially for the specific applications, but I think what I am going
%to say applies, more generally, to other   
%Physics fields as well as to other scientific disciplines.} 
(HEP) is well known for using 
very sophisticated detectors, 
status of the art computers, ultra-fast data acquisition systems,
and very detailed (Monte Carlo) simulations. One might imagine 
that a similar level of refinement 
could be found in its analysis tools, on the
trail of 
the progress in probability theory and statistics of the 
past half century. Quite the contrary! As pointed out by 
the downhearted Zech\cite{Zech}, 
``some decades ago physicists were usually well educated 
in basic statistics  in contrast to their colleagues in social and 
medical sciences. Today the situation is almost reversed. 
Very sophisticated methods are used in these disciplines, whereas 
in particle physics standard analysis tools available in many program
packages seem to make knowledge of statistics obsolete. This leads 
to strange habits, like the determination of a r.m.s. of a sample 
through a fit to a Gaussian. More severe are a widely spread 
ignorance about the (lack of) significance of $\chi^2$ 
tests with a large number of bins and missing experience 
with unfolding methods". 
In my opinion, the main reason 
for this cultural gap is that    
statistics and probability are not given 
sufficient importance in the 
student curricula: {\it ad hoc} formulae are provided in laboratory
courses to 
report the ``errors'' of measurements; 
the few regular lectures 
on ``statistics'' usually mix up  
{\it descriptive} statistics, {\it probability} theory and 
{\it inferential} statistics. 
This leaves a lot of freedom for personal interpretations of the 
subject (nothing to do with subjective probability!). 
Of equal significance is the fact that  
the {\it disturbing catalog of inconsistencies}\cite{Efron} of 
``conventional''\footnote{I prefer to call the frequentist approach  
``conventional'' rather than ``classical''.} 
 statistics
helps to give the impression that this subject is  matter of 
initiates and local gurus, rather than a scientific 
discipline\footnote{Who can have failed to experience
endless discussions about trivial statistical problems, the solution 
of which was finally accepted just because of the (scientific or, more
often, political) authority of somebody, rather than 
because of the strength of the logical arguments?}. 
 The result is that standard knowledge of statistics at 
the end of the University curriculum 
is insufficient and confused, 
as widely recognized.  
Typical effects of this (dis-)education are the 
{\it ``Gaussian syndrome''}\footnote{I know a senior physicist
who used to teach students that 
standard deviation is meaningful only for the Gaussian, and that 
it is ``defined'' as half of 
the interval around the average which contains 68\,\% of the events! 
More common is the evaluation of the standard deviation of a data sample
fitting a Gaussian to the data histogram
(see also the previous quotation of Zech\cite{Zech}), even
in those cases in which the histogram has nothing to do with 
a Gaussian. The most absurd 
case  I have heard of is that of someone 
fitting a Gaussian to a histogram exhibiting a flat shape
(and having good reasons for being considered to come from coming from 
a uniform distribution) 
to find the resolution of a silicon strip detector!.} (from which 
follows the uncritical use of the rule of combining results, 
weighing them with inverse of the ``error'' 
squared\footnote{For instance, the 1998 issue of the Review
of Particle Physics\cite{PDG} includes an example based on this kind 
of mistake with the intention to
show that ``the Bayesian methodology \ldots is not 
appropriate for the objective presentation of experimental data'' 
(section 29.6.2, pag. 175).}, 
or the habit of calling $\chi^2$
any sum of squared differences between fitted curves and data points,
and to use it as if it  were a $\chi^2$),
the abused {\it ``$\sqrt{n}$  rule''} to evaluate 
``errors'' of counting experiments\footnote{Who has never 
come across somebody
calculating the ``error'' on the efficiency $\epsilon = n/N$, using 
the standard ``error propagation'' starting from $\sqrt{n}$
and $\sqrt{N}$?} and the reluctance to take into account 
correlations\footnote{In the past, 
the correlation matrix was for many HEP physicists
``that mysterious list of numbers 
printed out by MINUIT'' (MINUIT\cite{MINUIT} 
is the minimization/fitting package mostly 
used in HEP), but at least some cared
about  understanding what those numbers meant and how to use them
 in further analysis. 
Currently - I agree with Zech\cite{Zech} - the situation has worsened: 
although many people do take into account correlations,
especially in combined analysis of crucial Standard Model parameters, 
the interactive packages, which display only 
standard deviations of fitted parameters, tend to ``hide'' the 
existence of correlations 
to average users.}, just to remain at a very basic level. 

I don't think that researchers in medical science or in biology 
have a better statistics education than physicists. On the contrary, 
their usually scant knowledge of the subject forces them to 
collaborate with professional statisticians, and this is the reason
why statistics journals contain plenty of papers in which sophisticated
methods are developed to solve complicated problems in the 
aforementioned fields. Physicists, especially in HEP, tend to be 
more autonomous, because of their 
skills in mathematics and computing, plus of a good dose of intuition.
But one has to admit that
it is rather unlikely that a physicist, in a constant 
hurry to publish results before anybody else, 
can reinvent methods which have been reached by others
 after years of 
work and discussion. 
Even those physicists who are considered experts in statistics 
usually read books and papers written and refereed 
by other physicists. The HEP community remains, therefore, isolated 
with respect to the 
mainstream of research in probability and statistics. 

In this paper I will not try to review all possible 
methods used in HEP, nor to make 
a detailed comparison between
conventional and
Bayesian solutions to solve the same problems.  
Those interested in this kind of statistical and historical study 
are recommended to look at the HEP databases and 
electronic archives\cite{archives}.
% like
%SPIRES\cite{SPIRES}, the 
%Los Alamos electronic archive\cite{XXX}, the CERN preprint 
%server\cite{alice}, or HEPDOC\cite{hepdoc}. 
I think that the participants
in this workshop are more interested in learning about 
the attitude of HEP physicists towards the fundamental aspects 
of probability, in which framework they make uncertainty statements, 
how subjective probability is perceived, and so on.
The intention here will be,
finally,  to contribute to the debate around the 
question {\it ``Why isn't everybody
a Bayesian''}\cite{Efron}, recently turned into 
{\it ``Why isn't every physicist a Bayesian''}\cite{Cousins}.
The theses which I will try to defend are: 
\begin{itemize}
\item
there is a contradiction between  a
cultural background in statistics 
and the good sense of physicists, 
and physicists' intuition is closer to the 
Bayesian approach than one might na\"\i vely think;
\item
there are 
cases in which good sense alone is not enough and 
serious mistakes can be made; it is then that 
the philosophical and
practical advantages offered by the Bayesian approach become 
of crucial importance;
\item
there is a chance that the Bayesian approach can become widely 
accepted, if it is presented in a way which is 
close to physicists intuitions
and can solve the ``existential'' problem of 
reconciling two aspects which seem irreconcilable: 
subjective probability and the 
honest ideal of objectivity which scientists have.
\end{itemize}

Some of the points sketched quickly in this paper are discussed 
in detail 
in lecture notes\cite{dagocern} based on several seminars 
and minicourses given over the past years. These notes also contain 
plenty of general and HEP inspired applications.  

\section{Measurement Uncertainty}
An idea well rooted among physicists, especially nuclear and particle 
physicists, 
is that the result of a measurement must be reported 
with a corresponding {\it uncertainty}. What makes the 
measured values subject to a degree of uncertainty 
is, it is commonly said, 
the effect of unavoidable measurement {\it errors},
usually classified as {\it random} (or {\it statistical}) and 
{\it systematic}\footnote{This last statement may 
sound like a tautology, since ``error'' and 
``uncertainty'' are often used as synonyms. This 
hints to the fact that in this subject there is 
neither uniformity of language, nor of methods, as is recognized
by the metrological organizations, which have made 
great efforts to bring some order into the 
field\cite{DIN,BIPM,ISO,ISOD,NIST}. 
In particular, the International Organization for Standardization
(ISO) has published {\it ``Guide to the expression of uncertainty
in measurement''}\cite{ISO}, containing definitions, recommendations
and practical examples. For example, {\it error} is defined
as ``the result of a measurement minus a true value of the measurand''
{\it uncertainty} ``a parameter, associated with the result of a measurement,
 that characterize the dispersion of the values that could reasonably be 
attributed to measurand'', and, finally, {\it true value} ``a value
compatible with the definition of the given particular quantity''. 
One can easily see that it is not just a question of practical
definitions. It seems to me that
there is a well-thought-out philosophical
choice behind these definitions, 
although it is not discussed extensively in the {\it Guide}. 
Two issues
in the {\it Guide} that I find of particular importance are the 
discussion on the sources of uncertainty and the admission
that  all contributions to the 
uncertainty are of a probabilistic nature. 
The latter is strictly related to the 
subjective interpretation of probability, as 
admitted by the {\it Guide} and discussed in depth 
in \cite{dagocern}.  
(The reason why these comments on the ISO {\it Guide} have been
placed in this long footnote is that, unfortunately, 
the {\it Guide} is not yet
known in the HEP community and, therefore, has no influence
on the behaviour of HEP physicists about which I am going 
to comment here.
This is also the reason why I will often use in this paper 
typical expressions currently used in HEP and 
which are in disagreement with the ISO recommendations. 
But I will use these expressions preferably within quote marks,
like ``systematic error'' instead of ``uncertainty due to 
a recognized systematic error of unknown size''.)}. 

Uncertainties due to statistical errors are commonly 
treated using the 
frequentist  concept of confidence intervals, although the procedure
is so unnatural that the interpretation of the result is unconsciously
subjective (as will be shown in a while), 
and there are known cases (of great relevance in frontier
research) in which this approach is not applicable. 

As far as uncertainties due to systematics errors are concerned, 
there is no conventional consistent theory to handle them, 
as is also indirectly recognized by the ISO {\it Guide}\cite{ISO}. 
The ``fashion'' at the moment is to add them quadratically if they are
considered to be independent, or to build a covariance matrix 
if not. 
This procedure is not justified theoretically 
(in the frequentist approach)
and I think that it is used essentially because of
the reluctance of experimentalists 
to add linearly the dozens
of contributions of a complicated HEP measurement, as the old fashioned
``theory'' of maximum errors  
suggests doing\footnote{In fact, one can see that when there are only 2 or 3 
contributions to the ``systematic error'', there are still people who 
prefer to add them linearly.}.
The pragmatic justification for the 
quadratic combination of ``systematic errors'' is that one is using 
a rule (the famous ``error propagation'' 
formula\footnote{The most well-known version is that in which correlations are
neglected: \\
\begin{center}
$ \sigma^2(Y) = \sum_i\left(\frac{\partial Y}{\partial X_i}\right)^2
               \sigma^2(X_i)\,. $\\
\end{center} 
$Y$ stands for the quantity of interest, the value
of which depends
on directly measured quantities, 
calibration constants and other 
systematic effects (all terms generically indicated by $X_i$). 
This formula
comes from probability theory, 
but it is valid if $X_i$ and $Y$ are 
random variables, $\sigma(X_i)$ are standard deviation and 
the linearization is reasonable. 
It is very interesting to look
at  text books to see how this formula is derived. 
The formula is usually initially proofed referring to  random variables
associated to observables and then, suddenly, it is referred
to physics quantities, without any justification.})
which is considered to be valid 
at least for 
``statistical errors''. But, in reality, 
this too is not correct. 
 The use of this formula is again arbitrary in the case
of ``statistical errors'', if these have been 
evaluated from confidence intervals\footnote{As 
far as ``systematic errors'' are concerned the situation 
is much more problematic because the ``errors'' are not even 
operationally well defined: they may correspond to 
subjectivist standard deviations (what I consider to be correct,
and what corresponds to the ISO {\it type B} standard uncertainty\cite{ISO}),
but they can more easily be maximum deviations, $\pm 50$\,\% variation
on a selection cut, or the absolute difference obtained using two
assumptions for the systematic effect.}. 
In fact, 
there is no logical reason 
why a probabilistic procedure proved for standard deviations of
random variables (the observables) should also be valid 
for  68\,\% confidence intervals, which is considered, somehow, 
an uncertainty attributed to the true value.

These examples show quite well the 
contradiction between
the cultural background on probability and the practical good sense
of physicists. Thanks to this good sense, frequentist ideas are 
constantly violated, with the positive effect that at least  
some results 
are obtained\footnote{I am strongly convinced 
that a rigorous application of frequentist
ideas leads nowhere.}. It is interesting to notice 
that in simple routine applications 
these results are   
very close, both in value and in meaning, to those 
achievable starting from what I consider to be the 
correct point of view for handling 
uncertainty (subjective probability). There are, on the other hand, 
critical cases in which
scientific conclusions may be seriously mistaken. Before 
discussing these cases, let us look more closely 
at the terms of the claimed 
contradiction.

\section{Professed Frequentism versus Practiced Subjectivism}
\subsection{HEP physicists ``are frequentist''}
If one asks HEP physicists ``what is probability?'', one  will
realize immediately that they ``think they are'' frequentist.  
The same impression is got looking at the books and 
lecture notes they use\cite{libri}. 
Particularly significant, to get an overview of ideas and methods 
commonly used, 
are the PDG\cite{PDG}
and other booklets\cite{Formulae,Briefbook} which   
have a kind of explicit (e.g. \cite{PDG,Formulae}) 
or implicit (e.g. \cite{Briefbook})
{\it imprimatur} of HEP organizations.
 
If, instead,  one asks physicists what they think about probability
as {\it ``degree of belief''} the reaction is negative and 
can even be violent: ``science must be objective: there is no room 
for belief'', or ``I don't believe something. I assess it. This is
no a matter for religion!''. 

\subsection{HEP physicists ``are Bayesian''}\label{ss:beliefs}
On the other hand, 
if one requires physicists to express their opinion about 
practical situations in condition of uncertainty, 
instead of just standard examination questions, 
one gets a completely different impression. One realize vividly
that Science is indeed based on beliefs, 
very solid and well grounded beliefs, 
but they remain beliefs 
``\ldots in instrument types, in programs of experiment
enquiry,
in the trained, individual judgements about every local behavior 
of pieces of apparatus''\cite{Galison}.

Physicists find it absolutely natural to talk 
about the probability of hypotheses, a concept for which there is no
room in the frequentist approach. Also the intuitive 
way with which they figure out the result is, in fact, a 
probabilistic assessment on the true value. 
Try to ask what is the probability that the top quark mass is
between 170 and 180 GeV. No one\footnote{Certainly one may find people
aware of the ``sophistication'' of the frequentist approach, 
but these kinds of probabilistic statements are 
currently heard in conferences and no 
frequentist guru stands up to complain that the speaker is talking
nonsense.} 
will reply that the question has no sense, since ``the top quark 
mass is a {\it constant of unknown value}'' (as an orthodox frequentist
should complain). 
They will simply answer that the probability is such and such percent,
using the published value and ``error''.
They are usually surprised if somebody tries to explain to 
them that they ``are not allowed'' to speak of
probability of a true value. 

Another word which physicists find scandalous is ``prior'' 
(``I don't want to be influenced by prejudices'' 
is the usual reply). 
But in reality priors play a very important role in  laboratory
routines, as well as at the moment of deciding that a paper is ready for 
publication. They allow experienced physicists to realize that 
something is going wrong, that a student has most probably made a serious
mistake, that the result has not yet been corrected by all systematic
effects, and so on. Unavoidably, priors generate 
some subtle cross correlations among results, and there are well known cases 
of the values of physics quantities slowly drifting from an initial 
point, with all subsequent results being included in the ``error bar''
of the previous experiment. But I think that 
there no one and nothing is to blame
for the fact 
that these things happen (unless made on purpose): 
a strong evidence is needed 
before the scientific
community radically changes its mind, 
and such evidence is often achieved
after a long series of experiments. Moreover,
 very subtle systematic effects
may affect the data, and it is not a simple task for an experimentalist
to decide when all corrections have been applied, if he has no idea 
what the result should be.

\subsection{Intuitive application of Bayes' theorem}
There is an example which I like to give, in order to 
demonstrate that the intuitive reasoning  
which unconsciously transforms confidence intervals into
probability intervals for the true value is, in fact, very close
to the Bayes' theorem. Let us imagine we see a hunting dog in a forest 
and have to guess where the hunter is, knowing that there is a
50\,\% probability that the dog is within 100 m around him.
The terms of the analogy 
with respect to observable and true value are obvious. 
Everybody will answer immediately that, 
with 50\,\% probability, 
the hunter is within 100 m from the dog. But everybody will also agree 
that the solution relies on some implicit assumptions: 
uniform {\it prior} 
distribution (of the hunter in the forest) and  
symmetric {\it likelihood} (the dog has no preferred direction, 
as far as we know, when he runs away from the hunter). 
Any variation in the assumptions leads to a different 
solution. And this is also easily recognized by physicists,
expecially HEP physicists, who are aware of situations in which 
the prior is not flat (like the cases 
of a bremsstrahlung photon or of a cosmic ray spectrum)
or the likelihood is not symmetric (not all detectors have a nice 
Gaussian response).  
In these situations intuition may still help a qualitatively guess
to be made about 
the direction of the effect on the value of the measurand, 
but a formal application of the Bayesian ideas 
becomes crucial
in order to state a result which is consistent with what 
can be honestly learned from data.  

\subsection{Bayes versus Monte Carlo}\label{ss:ML}
The fact that Bayesian inference is not currently used in HEP 
does not imply 
that non-trivial inverse problems remain unsolved, or that
results are usually wrong. The solution often relies on extensive 
use of Monte Carlo (MC) 
simulation\footnote{If there is something in which HEP physicists
really believe, it is 
Monte Carlo simulation! It plays a crucial role in all analyses, 
but sometimes its use as a multipurpose brute force problem solver
is really unjustified and it can, from a cultural point of view, 
be counterproductive. For example, 
I have seen it applied to solve elementary
problems which could be solved analytically, like ``proving'' that 
the variance of the sum of two random numbers is the sum of the variances.
I once  found a sentence at the end of the solution
of a standard probability problem which I consider to
be symptomatic of this brute force behaviour:   
``if you don't trust logic, then you can make a little 
Monte Carlo\ldots''.} and on intuition. 
The {\it inverse} problem is then 
treated as a {\it direct} one. The quantities of interest are 
considered as MC parameters, and    
are varied until the best statistical agreement between simulation 
output and experimental data is achieved. In principle, this is 
a simple numerical implementation of Maximum Likelihood, but in reality
the prior distribution is also taken into account in the simulation 
when it is known to be non uniform (like in the aforementioned 
example of a cosmic ray experiment). So, in reality what it is often 
maximized
is not the likelihood, but the Bayesian {\it posterior} 
(likelihood$\times$prior), and, as said before, the result 
is intuitively 
considered to be a probabilistic statement for the true value. 
So, also in this case, the results are close to those obtainable
by Bayesian inference, especially if the 
posterior is almost Gaussian (parabolic 
{\it negative log-likelihood}). 
Problems may occur, instead, when the ``not used'' prior  
is most likely not uniform, or when the posterior is  
very non-Gaussian. In the latter case the  difference between 
mode and average of the distribution, and the evaluation of 
the uncertainty 
from the ``$\Delta(\mbox{log-likelihood})=1/2$ rule'' can  
make quite a difference to the result. 
 
\section{Explicit use of Bayesian methods in HEP}
Besides the intuitive use of Bayesian reasoning, there 
are, in fact, some applications in which the Bayes' theorem
is explicitly applied. This happens when frequentist methods
``do not work'', i.e. they give 
manifestly absurd results, 
or in solving more complicated problems than just  
inferring the value of a quantity, like, for example, the deconvolution
of a spectrum (``unfolding''). Nevertheless, these methods are 
mostly used with a utilitarian spirit, without 
having really understood the meaning of subjective probability, or even 
remaining skeptical about it. They are used as one uses one of
the many frequentist ``ad hoc-eries''\footnote{For example, 
this was exactly the attitude 
which I had some years ago, 
when I wrote a 
{\it Bayesian unfolding} program\cite{unfolding}, and that 
of the large majority my 
colleagues who still use the program. Now, after having attended 
the 1998 Valencia Meeting on Bayesian Statistics, I have realized 
that this pragmatic frequentist-like use of Bayesian methods 
is rather common.}, after it has been ``proved'' 
that they work by MC 
simulation\footnote{I would like to point out that sometimes the 
conclusions derived from 
MC checks of Bayesian procedures may be 
misleading, as discussed in detail in \cite{dagocern}.}. 

Some of the cases in which the conventional methods do not work 
have even induced the PDG\cite{PDG} to present Bayesian methods. 
But, according to the PDG, a paper published this year\cite{FC} 
finally gives a frequentist solution to the problems, and this 
solution is recommended for publishing the results. Let us review
the situation citing directly \cite{FC}: ``Classical confidence intervals 
are the traditional way in which high energy physicists report errors 
on results of experiments.\ \ldots\ In recent years, there has been
 considerable dissatisfaction \ldots for upper confidence limits\ldots
This dissatisfaction led the PDG to describe procedures for Bayesian 
interval construction in the troublesome cases: Poisson processes
with background and Gaussian errors with a bounded physical region.
\ \ldots\  In this paper, we\,\ldots use\, (\ldots) to 
obtain a unified set of classical confidence
intervals for setting upper limits and quoting two-sided confidence
intervals. \ldots We then obtain confidence intervals which are 
never unphysical or empty. Thus they remove an original motivation 
for the description of Bayesian intervals by the PDG.''
In fact, the 1998 issue of the {\it Review
of Particle Physics} still exhibits
the Bayesian 
approach (with the typical misconceptions 
that frequentists have about it), but then it 
suggests two papers by frequentists\cite{Cousins,Efron} (``a balanced 
discussion''\cite{PDG}) to help 
practitioners to form their own idea on the subject, and, finally, 
it warmly recommends the new frequentist approach.
It is easy to imagine what the reaction 
of the average HEP physicist will be when confronted by 
the authority of the PDG, unaware 
that ``the PDG'' which rules analysis methods is in reality
constituted of
no more than one or two persons who recommend a paper written by 
their friends (as is clear 
from the references and the cross acknowledgements). One should also 
notice
that this  paper  claims important progress in statistics, 
but was in fact published in a physics journal (I wonder what 
the reaction of a referee of a 
statistics journal would have been\ldots). 

In conclusion, there is still a large gap between good sense 
and the dominating statistical culture. For this reason we must still be 
very careful in interpreting published results and in 
evaluating whether or not 
the conventional methods used lead to 
correct scientific conclusions ``by chance''. 
Some cases of misleading results will be described 
in the next section.

\section{Examples of Misleading Results Induced by Conventional Statistics}
It is well known that the frequentist denial of the 
concept of probability 
of hypotheses leads to 
misleading results in all cases in which the 
simple ``dog-hunter inversion paradigm''
is violated. This also happens in HEP. 

As already discussed, confidence levels are intuitively thought 
(and usually taught) as probabilities for the true values. 
I must recognize that many frequentist books do insist 
on the fact that the probability statement is not referred
to the true value. But then, when these books have to explain the
``real meaning'' of the result, they are forced to use ambiguous 
sentences which remain 
stamped in the memory of the reader much more than the 
frequentistically-correct twisted reasoning that they 
try to explain. 
For example Frodesen et al.\cite{libri} speak about
``the faith we attach to this statement'', as if 
``faith'' was not the same as degree of belief\ldots); 
Eadie et al.\cite{libri} introduce the argument saying that 
``we want to find {\it the range} \ldots which contains the true value 
$\theta_\circ$ with probability 
$\beta$''\footnote{I think that Aristoteles would have gotten mad if 
somebody had tried to convince him that the proposition
``the range contains $\theta_\circ$ with probability $\beta$'' 
does not imply 
``$\theta_\circ$ is in that range with probability $\beta$''.}; 
and so on. 

Similarly, significance levels are usually taken
as probability of the tested hypothesis. Also this 
non-orthodox interpretation is stimulated 
by sentences like
``in statistical context, the words {\it highly significant}
mean {\it proved beyond a reasonable doubt}''.  
It is also well known that the arbitrary translation of 
{\it $p$-values} into probability of the null hypothesis 
produces more severe mistakes than those concerning 
the use of confidence interval for uncertainty statements 
on  true values. 

Let us consider some real life examples 
of the misinterpretation of the two kinds just described. 

\subsection{Claims of new physics based on {$p$-values}}\label{ss:HERA}
You may have heard in past years some rumors, or even 
official claims, of discoveries of ``New Physics'', i.e. of   
phenomenology which goes behind the so called Standard Model
of elementary particles. Then, after some time, 
these announcements 
were systematically 
recognized as having been false alarms, 
with a consequent reduction in the 
credibility of the HEP 
community in the eyes 
of public opinion and tax payers (with easily imaginable 
long term aftermath for government  support of this research).
All these fake discoveries were 
based on considering low $p$-values as probability 
of the null hypothesis ``no new effect''. 
The most recent example of this kind 
is the so called 1997 ``HERA high $Q^2$ events excess''. 
The 
H1 and ZEUS collaborations, analyzing data collected at the HERA
very high energy electron-proton collider in Hamburg (Germany), found 
an excess of events (with respect to expectations) in the kinematical 
region corresponding to very hard interactions\cite{H1Z}. The 
``combined significance''\footnote{Physicists are not familiar 
with the term {\it $p$-value} (readers not familiar with this
term may find a concise review in 
\cite{pvalues}). Moreover, they are usually not aware of the implications
of the fact 
that the statistical significance takes into account 
also the probability of 
unobserved data (see, e.g., \cite{BB}).}
 of the excess was of the order of 1\,\%. 
Its interpretation as a hint of new physics was even suggested
by official statement by the laboratory and by other agencies. 
For example the DESY official statement 
was
{\it ``\ldots  the joint 
distribution has a probability of less than one percent to come
from Standard Model NC DIS processes''}\cite{HERA} 
(then it implies ``it has a $> 99$\,\% 
probability of not coming from the standard 
model!''\footnote{One might think that  
the misleading meaning of that sentence 
was due to unfortunate wording, 
but this 
possibility is ruled out by other statements which show
clearly a quite odd  point of view of probabilistic matter. 
In fact the DESY 1998 activity report\cite{DESYreport} insists 
in saying that 
``the likelihood that the data produced is the result
of a statistical fluctuation, \ldots, is equivalent to that 
of tossing a coin and throwing seven 'heads' or 'tails'
in a row'' (replacing 'probability' by 'likelihood' does
not change the sense of the message). 
Then, trying to explain the meaning of a 
statistical fluctuation, the following example is given: 
``This process can be simulated with a die. 
If the number of times a die is
thrown is sufficiently large, the die falls equally often on all faces, 
i.e. all six numbers occur equally often. The probability for
each face is exactly a sixth or 16.66\,\% - assuming the die 
is not loaded. If the die is thrown less often, then the probability
curve for the distribution of the six die values is no longer a straight
line but has peaks and troughs. The probability distribution 
obtained by throwing the die varies about the theoretical value
of 16.66\,\% depending on how many times it is thrown''. }). 
Similarly, the Italian INFN reported that {\it ``la
probabilit\`a che gli eventi osservati siano una fluttazione
statistica \`e inferiore all'1\,\%''} (then, 
it implies that ``with 99\,\% probability, 
the events are not a statistical fluctuation, i.e. new physics''!). 
This is the reason why the press reported the news as 
``scientists are
practically sure they  have found new physics''. 
What I found astonishing is that most of the people I talked to 
had real difficulty in understanding that this probability inversion
is not legitimate. Only when I forced them to state
their degree of belief using the logic of the coherent bet
did it emerge that most of my colleagues would not even place a
1:1 bet in favour of the new discovery. Nevertheless, 
they were in favour of publishing the result because the loss 
function was absolutely unbalanced 
(an indirect Nobel prize against essentially nothing).

\subsection{What does a lower mass bound mean?}
The second example concerns 
confidence intervals, and it comes from new particle search. This has
always been one of the main activities of HEP.
New particles are postulated by theories and experimentalists
look for evidence for them in experimental data. Usually,
if the particle is not ``observed''\footnote{This concept of ``observation''
is not like that of seeing a black swan, 
to mention a famous classical example. New particles leave 
signatures in the detector that on an event by event basis 
cannot be distinguished by other processes (background). 
A statistical (inferential) analysis is therefore needed.} 
one says that, although 
the observation does not disprove the existence of the particle,
this is an indication of the fact that the 
particle is ``too heavy''. 
The result is then quantified by a ``lower bound'' at a
``95\,\% confidence level''. Without entering into detail
of how the limit is operationally defined (see, e.g., \cite{LEP4} and 
references therein, in particular 
\cite{JL}, to have an idea of the level 
of complication 
reachable to solve a simple problem), 
I want to point out that also in this case the
result can be misleading. Again I will give a real life
example. A combined analysis of all the 
LEP experiments on the Higgs mass concluded recently that
{\it ``A 95\,\% confidence level lower bound of 77.5 GeV/c$^2$ is 
obtained for the mass of the Standard Model Higgs 
boson''}\footnote{In the meanwhile new data have increased this limit,
but the actual number is irrelevant for this discussion.}. 
This {\it sounds as if} one were sure at 95\,\% that the mass is above
the quoted bound. In fact, most of the people I 
interviewed about the meaning of the statement, 
even those belonging to
the LEP experimental teams, answered  ``\underline{if} the Higgs boson 
exists at all, then there is 95\,\% probability 
that its mass is above the limit''. There were also a few people 
who answered  
``if I do a MC simulation of the decay of a 77.5 GeV Higgs
boson, I get in only 5\,\% of the cases the simulation describing the data'',
or ``if there is an Higgs boson and its mass is less than 77.5 GeV,
then the observations of the search experiments have a probability of 
at most 5\,\% of being correct'', or something similar. 
From all of which it is
very difficult to understand, from a logical point of view, why 
one should be 95\,\% confident that the mass is higher than the 
bound\footnote{There was also somebody who refused to answer 
because ``your question is going to be difficult to answer'', or 
without any justification (perhaps they 
realized that it was impossible to explain the statement 
to a scientific journalist, or to a government authority - these were
the terms of my question - without using probabilistic statements 
which were incompatible with what they thought about probability).}.

The problem can be solved easily with Bayesian methods
(see \cite{dagocern} for details).
Assuming a flat prior for the mass, one 
finds that the value of the lower bound is more or less
the published one, but only under the condition that 
the mass does not exceed the kinematical limit of the studied reaction. 
But this limit is just a few GeV above the stated lower bound. 
Thus in order to obtain the correct result one needs to renormalize
the probability taking account of the possible range of masses above the 
kinematical limit and for which the 
experiment has no sensitivity. For this reason, in the case of \cite{LEP4}
the probability that 
the mass value is above 77.5 GeV/c$^2$ may easily become 99.9\,\%, 
or more, depending on the order of magnitude of a possible upper 
bound for the mass.
Then,  in practice these lower bounds can be taken as 
certainties\footnote{There are in fact theorists who ``assume''
the lower bounds as certain bounds in their considerations. 
Perhaps they do it intuitively, or because they have
heard in the last decades of thousands of these 95\,\% lower bounds, 
and never a particle has then shown up in the 5\,\% side\ldots}.

\section{Conclusions}
Although it is clear that the dominant statistical culture is still
frequentism in HEP (and everywhere else),
I am myself rather 
optimist on the possibility that 
the situation will change, at least in HEP, and that Bayesian
reasoning will emerge from an intuitive to a 
conscious level. This is not a dream (although clearly several academic 
generations are still needed) if the theory is 
presented in a way that it is acceptable to an ``experienced 
physicist''. 
\begin{itemize}
\item
First, it is not difficult to get a consensus 
on the observation that 
subjective probability is the natural concept developed by
the human mind to quantify the plausibility of events in 
conditions of uncertainty. 
\item
Second, one should insist on the fact 
that Bayes' theorem is in fact 
a natural way of reasoning in updating probability, 
and \underline{not} a philosophical point of view 
that somebody tries to 
apply to data analysis\footnote{For example
frequentists completely
misunderstand this points, when they state, e.g., 
that ``Bayesian methods proceed by invoking an interpretation of
Bayes' theorem, in which one deems it sensible to consider a p.d.f. 
for the unknown true value $m_t$'', or 
that ``a pragmatist can consider the utility of equations 
generated by the two approaches while skirting the issue of 
buying a whole philosophy of science''\cite{Cousins}.

I find that  also the Zellner's paper\cite{Zellner} demonstrating 
that Bayes' theorem makes the best use of the 
available information can 
help a lot to convince people.} (see \cite{dagocern}
for details). 

\item
Bayes' theorem is not ``all''. It only works in situations 
where the nice scheme of prior and likelihood is applicable. 
In many circumstances one can assess a subjective probability
directly  
(try asking a carpenter how much he believes the result of
 his measurement!). 
\item
The {\it coherent bet} (\`a la de Finetti\cite{definetti}) forces people 
to be honest and to make the best (i.e. ``most objective'') 
assessments of probability.
\item
It is preferable not to mix up probability evaluation with 
decision problems\footnote{Although it may seem absurd, 
the Bayesian approach is recognized by ``frequentists'' 
to be ``well adapted to
decision-making situations''\cite{PDG}(see also \cite{Cousins,FC}).
I wonder what then probability is for these authors.}. 
In other words, the coherent bet 
should be considered  hypothetical. This makes a clear distinction 
between our beliefs and our wishes (the example in section \ref{ss:HERA} 
should teach something in this respect).  
\item
One may think, na\"\i vely, 
that the ``objective Bayesian theory'' is more 
suited for science than the 
``subjective one''. Instead, it seems to me  easier to convince     
experienced physicists that ``there is nothing really that is
objective'', 
than it is to accept an objective theory containing 
ingredients which appear 
dogmatic\footnote{Dogmatism is never desirable. It can be easily 
turned against the theory. 
For example, one  criticism of \cite{FC} says, more or less, that 
Bayesian theory supports Jeffreys' priors, and not uniform priors,
but, since Jeffreys' priors give unreasonable results in their
application, then one should mistrust Bayesian methods! 
(see also \cite{valencia6}.) One may object that the
meaning and the role of Jeffreys' priors was misunderstood, 
but it seems to me  difficult to control the use of {\it objective priors} 
or of {\it reference analysis} once they have left the community
of experts aware of the ``rather special nature and role 
of the concept of a `minimally informative' prior specification
- appropriately defined!''\cite{BS}.} 
from the physicist point of view\cite{valencia6,dagocern}. 
Any experienced physicist knows already that the only 
``objective'' thing 
in science is the reading of digital scales. When we want to
transform this information into scientific knowledge we have to make 
use of many implicit and explicit beliefs
(see section \ref{ss:beliefs}).
Nevertheless, 
the ``honest'' (but na\"\i ve) 
ideal of objectivity can be recovered if  
scientific knowledge is considered  
 as a kind of very {\it solid networks of beliefs}, 
based on centuries of experimentation, with {\it fuzzy borders} 
which correspond to the areas of present research. 
My preferred motto is that ``no one should be allowed to talk about 
objectivity, unless he has 10 or 20 years of experience in frontier
science, economics, or any other real world application''. In 
particular, mathematicians should refrain from using the word 
objectivity when talking about the physical world.  
\item
It is very important to work on applications: 
the simplicity and the naturalness of the Bayesian reasoning 
will certainly attract people. 
\item
Many conventional methods can be easily recovered as limit cases 
of the Bayesian ones, if some well defined 
restricting conditions are valid, as already 
discussed in section \ref{ss:ML}. For example, when I make 
a $\chi^2$ fit
I consider myself to be using a Bayesian method, 
although in a simplified form. This attitude contrasts 
to that of practitioners who use methods in which the Bayes' theorem 
is explicitly applied, but as if it were one of the many 
frequentist cooking recipes. 
\item
It is important to make efforts to introduce Bayesian thinking
in teaching, starting from the basic 
courses. I am not the first to have realized that the Bayesian approach 
is simple for students. The resistance comes from our colleagues, 
who are unwilling to renew the contents of their lectures, and who have 
developed a distorted way of thinking.
\end{itemize}
Finally, I would like to give a last recommendation. Don't try 
to convince a physicist that he already is Bayesian, or that 
you want to convert him to become Bayesian. 
A physicist feels offended if you call
him ``X-ian'', be it Newtonian, Fermian, or Einsteinian.
But, being human, he has a natural feel for
probability, just like everybody else. 
I would like to generalize this idea and propose 
reducing the use of the adjective
``Bayesian''. I think that the important thing is to have 
a theory of uncertainty in which ``probability'' has the same meaning 
for everybody, precisely that meaning which the human mind has 
naturally developed and that frequentists have tried to kill.  
Therefore I would rather call these methods {\it probabilistic 
methods}. 
And I conclude saying that, obviously, 
``\underline{I am not a Bayesian}''.

\section{Acknowledgements}
It is a pleasure to thank the organizers of Maxent98 
for the warm hospitality in Garching, which  favoured 
friendly and fruitful discussions among the participants.

\end{document}